\documentclass{article}
\usepackage{spconf,amsmath,graphicx}
\usepackage{makecell}
\usepackage{multirow}

\title{adversarial attack driven data augmentation for accurate and robust medical image segmentation}
\name{Mst. Tasnim Pervin$^{\star}$\qquad Linmi Tao$^{\star}$ \qquad Aminul Huq$^{\star}$\qquad Zuoxiang He$^{\dagger}$ \qquad Li Huo$^{\ddagger}$
\thanks{Correspondences should be addressed to : linmi@tsinghua.edu.cn. This work is supported by the National Science Foundation of China at the Project 61672017.}}
\address{$^{\star}$Department of Computer Science and Technology, Tsinghua University, Beijing, China\\
$^{\star}$Beijing National Research Center for Information Science and Technology, China\\
$^{\star}$Key Laboratory of Pervasive Computing, Ministry of Education, Beijing, China\\
$^{\dagger}$Beijing Tsinghua Changgung Hospital, School of Clinical Medicine, Tsinghua University\\
$^{\ddagger}$M.D, Nuclear Medicine Department, Peking Union Medical College Hospital
}

\begin{document}
\topmargin=0mm
%
\maketitle
\begin{abstract}
Segmentation is considered to be a very crucial task in medical image analysis. This task has been easier since deep learning models have taken over with its high performing behavior. However, deep learning model’s dependency on large data proves it to be an obstacle in medical image analysis because of insufficient data samples. Several data augmentation techniques have been used to mitigate this problem. We propose a new augmentation method by introducing adversarial learning attack techniques, specifically Fast Gradient Sign Method (FGSM). Furthermore, We have also introduced the concept of Inverse FGSM (InvFGSM), which works in the opposite manner of FGSM for the data augmentation. This two approaches worked together to improve the segmentation accuracy, as well as helped the model to gain robustness against adversarial attacks. The overall analysis of experiments indicates a novel use of adversarial machine learning along with robustness enhancement.
\end{abstract}
\section{Introduction}
\label{sec:intro}
The substantial structures of Deep Learning (DL) models have been proved to be very effective for several discriminating tasks related to Computer Vision applications such as prediction, image segmentation, or object recognition with the help of progressive and efficient computational power, and in-depth network architecture 
\cite{shorten2019}. However, DL networks can rarely perform well without the availability of a massive dataset. This limitation proves to be even larger for the field of medical image processing as the availability of a big data is not an easy coin to toss. To mitigate these obstacles, several supervised biomedical segmentation methods focus on manually-engineered preprocessing steps and architectures \cite{moeskops2016}, \cite{pereira2016}. Manually-tuned data augmentation is also used to increase the number of training examples 
\cite{marques2017}, \cite{roth2015}.

Data augmentation is a good solution for this problem as it will minimize the gap between validation and training portions. Augmentation functions such as random rotations or random nonlinear deformations are easy to apply and are fruitful at improving segmentation accuracy in some settings \cite{moeskops2016}, \cite{pereira2016}, \cite{roth2015}. However, these functions have a limited capability to imitate real variations and can be highly subtle to the choice of parameters \cite{dosovi2015}. This phenomena has been reported in the paper of skin lesion classification by \cite{esteva2017} and so for liver lesion classification \cite{litjens2017}. 

In this paper, we carried out the task of data augmentation by employing adversarial learning. Recent works show that adversarial attacks are equally capable to degrade segmentation models' performance \cite{paschali2018}. The traditional augmentation techniques can not make these models' robust against adversarial attacks. Wetstein et al. proved that powerful DL models such as Inception-v3 and DenseNet-121 tend to loose strength when attacked by FGSM and Projected Gradient Descent (PGD) attacks \cite{wetstein2020}. Paschali et al. also investigated the effect of adversarial attack in degrading performance of classification (Inception V3, Inception V4, MobileNet) and segmentation (SegNet, U-Net, DenseNet) models \cite{paschali2018}. In a recent work, the authors Chen et al. used adversarial data augmentation for improved generalization ability by generating natural signal corruptions rather than pixel attacks \cite{chen2020}. Though adversarial attacks are meant to degrade ML/ DL model performance, but here we used an adversarial attack to enlarge the dataset for training segmentation model. We have also introduced a new perspective to use FGSM for further improvement by modifying its working technique. 
The key contribution of this paper can be listed as below:
\begin{itemize}
	    \item We used adversarial learning in favor of deep learning models rather that traditional aim of making models' vulnerable. We used FGSM attack to create adversarial examples (tuned to have minimum noise). This helped to double the number of data to be augmented to mitigate overfitting.
	    \item We used inverse approach of FGSM (InvFGSM) which targets to adjust input data towards minimum loss. This helped model with improved performance by adding positive noises instead of adversarial noises. This approach has not been used till now in any other literature.
	    \item Generally all DL models are vulnerable to adversarial attacks unless are adversarial trained or use any defense mechanism. For our work, we used adversarial training based augmentation, so the model became prepared and become robust against similar attacks in future.
	\end{itemize}

\section{Related Works}
Medical image segmentation models uses similar architecture with fewer convolutional blocks and fewer parameters. Ronnerberger et al. proposed the most well-known segmentation network, U-Net, where they expanded the FCN into upsampling and downsampling layers to work with few training images and to create high accurate output \cite{ronnerberger2015}. Since then, many modification of U-Net has been proposed to improve the segmentation in various applications, for instance \cite{novikov2018}. Yuan created a network called Convolutional-Deconvolutional Neural Networks that uses Jaccard distance based loss function and performs under different image acquisition conditions \cite{yuan2017}. For consistent use of contextual information, Hwang el al. proposed a model called Network-Wise Training of Convolutional Networks that employs a multi-stage training strategy for distilled segmentation with smooth boundary \cite{hwang2017}. 

Recent advances in adversarial learning has broken the phantasm of high-performing DL models. An adversarial example, created using attacks of adversarial learning, is a sample of modified inputs (image, text, audio, etc.) that can force a model to misbehave. These attacks corrupt DL models by injecting tiny amount of imperceptible noise into the input and generates original input look-alike adversarial examples. These examples pose a potential security threats for practical applications of DL. Engstrom et al. showed that deep CNN models can be forced to do misclassification by attacking with some easy geometric transformations. This phenomena of performance degradation has been noticed for several dataset including MNIST, CIFAR10, and ImageNet (26\% drop for MNIST, 72\% drop for CIFAR10, and 28\% drop for ImageNet)\cite{engstrom2017}. In another research, Ian Goodfellow and his fellow authors introduced to Fast Gradient Sign Method using a maxout network to generate adversarial examples which produces 89.4\% misclassification with 97.6\% confidence \cite{goodfellow2014}. Li et al. in their work compared the performance of original testing data with a new type of adversarial examples \cite{li2018}. Su et al. invented one pixel attack where they have shown that changing a single pixel of input images can cause 70.97\% of misclassification \cite{su2019}. Zajac and his fellow researchers have focused on attacking the border of images \cite{zajac2019}. These examples, and many others, have demonstrated DL models are not robust to the attacks, which is harmful to the applications especially in health area. 
 
 \begin{figure}[t]
    \centering
    \includegraphics[width=8.8cm, height = 6cm]{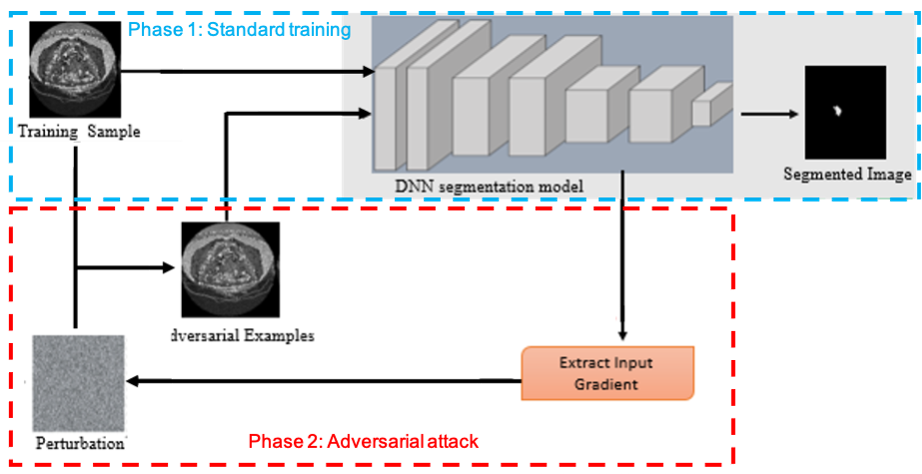}
    \caption{Workflow of proposed approach}
    \label{fig:Workflow of proposed approach}
\end{figure}
 
\section{Methodology}
The proposed methodology poses a heavy contrast to the traditional augmentation techniques. The effectiveness of adversarial training in the form of noise or augmentation search is a new concept that has not been widely tested and understood. The flow diagram of figure 1 aims to represent the whole procedure: 1) Train model with original dataset for segmented image localization, 2) Attack the model with adversarial attack technique that will find model architecture and input gradients to craft noises and create new adversarial examples, 3) Adversarial examples will be augmented to the original samples to create new training set that will be fed to the model for training which can produce better segmentation result and make model robust against attacks. The segmentation model and attack techniques has been described briefly in following sections.
\subsection{U-Net}
\begin{figure}[t]
    \centering
    \includegraphics[width = 7.5cm, height =1.9cm]{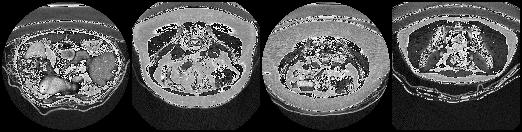}\\
    (a)\\
    \includegraphics[width = 7.5cm, height =1.9cm]{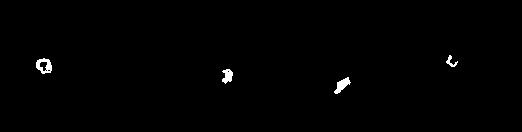} \\
    (b)\\
    \includegraphics[width = 7.5cm, height =1.9cm]{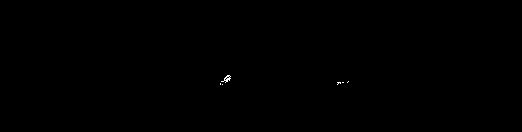}\\
    (c)
    \caption{Performance of UNet without any augmentation : (a) Original input, (b) Ground truth/mask, (c) Predicted mask}
    \label{fig:my_label}
\end{figure}
The U-Net model, a semantic segmentation model, was developed by Olaf Ronnerberger \cite{ronnerberger2015}. The model is built of a contraction/encoder path and an expanding/decoder path. Encoder path is nothing but a pile containing convolutional and maxpooling layers that emphasizes on the context or ‘What’ information of image where decoder path summons some transposed convolutional layers pursuing the location or the ‘Where’ information. In the encoder path, the size of the images starts to gradually decrease with increasing depth such as starting from $128\times128\times3$ to $8\times8\times256$. In the decoder path, the size of the image gradually increases with decreasing depth such as from $8\times8\times256$ to $128\times128\times1$. For a specific localization, a concatenation takes place at each level between the encoder's feature maps with the decoder's transposed convolution. This characteristic gives the architecture a U shape. The strength of U-Net is that the network only contains fully convolutional layers but no dense layer which enables model to accept images of any size as input.
\subsection{Fast Gradient Sign Method}
\begin{figure}[t]
\centering
    \includegraphics[width = 7.5cm, height =1.9cm]{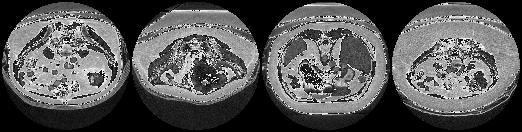}  \\
    (a) \\
    \includegraphics[width = 7.5cm, height =1.9cm]{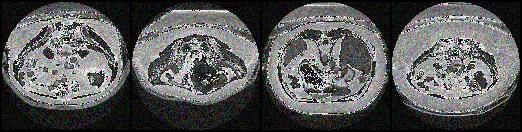} \\
    (b)\\
    \includegraphics[width = 7.5cm, height =1.9cm]{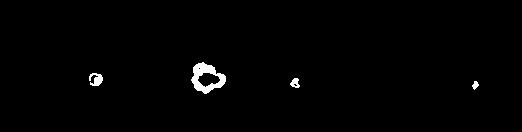} \\
    (c)\\
    \includegraphics[width = 7.5cm, height =1.9cm]{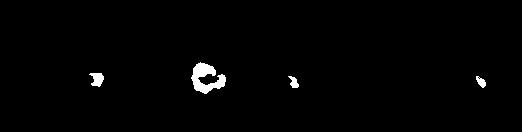} \\
    (d)
    \caption{Performance of UNet after adversarial augmentation by FGSM : (a) Original input, (b) Adversarial image for augmentation created by FGSM, (c) Ground truth/mask, (d)Predicted mask after augmentation }
\end{figure}
The most popular adversarial attack, Fast Gradient Sign Method (FGSM) has been proposed by Ian J. Goodfellow \cite{goodfellow2014}. The method works as follows: Suppose $x  \&  y$ are the input images and corresponding labels respectively, the weights of the network is represented by $\theta$ and the loss function is $L(\theta, x, y)$ . First, the gradient of the loss function for input image is calculated. The $\nabla$ operator takes out the derivative of the loss function. It gives a shape matrix including width, height, channels, and the tangent slopes. But most importantly only the sign of the slopes is needed to know when the pixel values increase or decrease. Then, these slope signs are multiplied by a very tiny imperceptible perturbation value $\epsilon$ . The resultant pattern is then added to the original inputs to create respective adversarial examples. The mathematical representation of this procedure is as follows:
\begin{equation}
    \centering
    x_{adv} = x + \epsilon * sign(\nabla_{x}L(\theta ,x,y))
\end{equation}
\subsection{Inverse Fast Gradient Sign Method }
The motivation for this approach came from the definition of FGSM. From the gradient of the loss function, we can assume how the changes occur in the input image. FGSM approach targets to adjust the input to a maximum loss using backpropagated gradients rather than adjusting weights to decrease the loss. Inverse FGSM (InvFGSM) adopts an inverse approach of adjusting input to a minimum loss using gradient which definitely helps to gain better performance. For adversarial noise in the FGSM approach, only the sign of the gradient matters. As the sign of the derivatives provides adversarial noises which are added with input to generate adversarial examples, the inverse of the sign provides new type of adversarial noise that will work positively aiding in loss minimization. Data augmentation with these kind of adversarial examples with positive noises seem to be useful in enhancing model performance. The mathematical way to represent this would be as follows:
\begin{equation}
    \centering
    x_{adv} = x - \epsilon * sign(\nabla_{x}L(\theta ,x,y))
\end{equation}

\section{Experimental Analysis}
\subsection{Dataset Description}
 For this experiment, we have used a dataset of colon cancer provided by Beijing Tsinghua Changgung Hospital and Peking Union Medical College Hospital. This dataset contains 1285 CT scan images along with same number of mask images of desired localization of segmentation. The original images are of size $512\times512\times3$  and resized into $128\times128\times3$ for faster processing when feeding into the model with the support of Jittor platform \cite{Jittor2020}.
 
\subsection{Data Augmentation by Adversarial Training}

\begin{figure}[t]
\centering
    \includegraphics[width = 7.5cm, height =1.9cm]{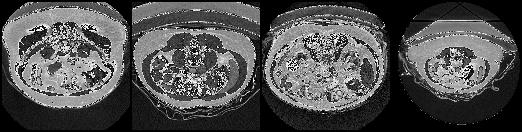} \\
    (a)\\
    \includegraphics[width = 7.5cm, height =1.8cm]{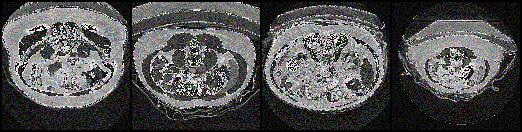}\\
    (b)\\
    \includegraphics[width = 7.5cm, height =1.8cm]{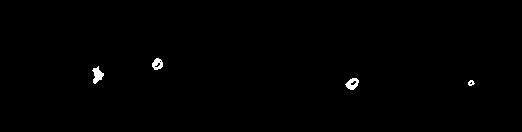}\\
    (c)\\
    \includegraphics[width = 7.5cm, height =1.8cm]{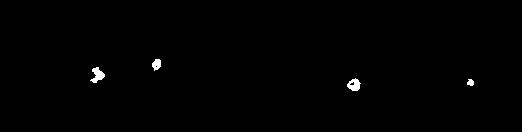}\\
    (d)
    \caption{ Performance of UNet after adversarial augmentation by inverse-FGSM : (a) original input, (b) adversarial image for augmentation created by inverse-FGSM, (c) Ground Truth/Mask, (d)predicted segmented image after augmentation }
\end{figure}

For experiment, the parameters of model training without data augmentation and with data augmentation have been set as: Loss Function = Categorical Cross-Entropy, Optimizer = SGD, Learning Rate = 0.1, Momentum = 0.99, Batch-size = 16. From table 1, the difference between the resultant Intersection over Union before and after augmentation can be noticed. The U-Net model performed quite well giving about 80\% correct segmentation till epoch 30. Data augmentation helped this performance even better. UNet along with FGSM performed approximately 3\% better than U-Net alone. But as suspected, Inverse FGSM gave a boost to this performance even further showing about 6\% raise. The figures 2,3,4 give indication to this phenomena as well.
Augmentation with adversarial machine learning attack techniques can help making model more robust. This phenomena has been shown in table 2. We employed a range of random perturbation starting from 0 to 0.2 to check the robustness of the model. It can be noticed how increasing noise affect the model. Even Noise of $\epsilon=0.2$ pulls down the segmentation performance up to 35\%.  Though adversarial training definitely helped the models to regain their performance and made the model robust. For adversarial training, the perturbation was kept to 0.1 as, visual imperceptibility is another aim along with good performance and higher rate of perturbation makes adversarial images visually corrupted. 
\begin{table}[t]
\centering
\caption{Performance of models over increasing epochs}
\vspace{0.2cm}
\begin{tabular}{m{2.5cm}|m{1cm}|m{1cm}|m{1cm}|m{1cm}}
\multirow{2}{*}{Models} & \multicolumn{4}{c}{IoU over epochs}               \\\cline{2-5} \vspace{3cm}
                        & $0^{th}$ & $10^{th}$ & $20^{th}$ & $30^{th}$ \\ \hline
UNet                    & 0.4955   & 0.4986   & 0.4983   & 0.8084    \\
UNet + FGSM             & 0.4987   & 0.7839   & 0.8129   & 0.8394    \\
UNet + Inverse- FGSM    & $\boldsymbol{0.6159}$   & $\boldsymbol{0.8264}$   & $\boldsymbol{0.8643}$   & $\boldsymbol{0.8986}$  
\end{tabular}
\end{table}

\begin{table}[t]
\centering
\caption{ Effect of attack on model robustness with increasing perturbation($\epsilon$)}
\vspace{0.2cm}
\begin{tabular}{m{1.2cm}|m{1.5cm}|m{1cm}|m{1cm}|m{0.8cm}}
\begin{tabular}[c]{@{}c@{}}Attacking \\ model\end{tabular} &
  \begin{tabular}[c]{@{}c@{}}Adversarial \\ training\end{tabular} &
  $\epsilon =  0.0012$ &
  $\epsilon  = 0.118$ &
  $\epsilon  =  0.1176$ \\ \hline
\multirow{2}{*}{FGSM}         & before & 0.7860 & 0.7246 &  0.6028\\
                              & after  & $\boldsymbol{0.9071}$ & $\boldsymbol{0.8242}$ & $\boldsymbol{0.7720}$ \\ \hline
\multirow{2}{*}{\begin{tabular}[c]{@{}c@{}}Inverse-\\ FGSM\end{tabular}} & before & 0.7749 & 0.5841 & 0.5467 \\
                              & after  & $\boldsymbol{0.8724}$ &$\boldsymbol{0.8729}$ & $\boldsymbol{0.8477}$
\end{tabular}
\end{table}

\section{Conclusion}
In this paper, we tried to apply a deep learning model for cancer cell segmentation from colon cancer dataset and also introduced a new approach for data augmentation. Limited data causes the overfitting problem when used on deep learning models, that’s why data augmentation is considered to be a crucial step. Though it cannot be an ultimate backup for limited data where any class sample is absent, it can help to deduct overfitting by generating a large number of similar data. We experimented with adversarial machine learning attack techniques as a new data augmentation approach that proved to be beneficial in improving the segmentation performance for this dataset. Though the conventional aim of adversarial machine learning is related to model robustness, this work certainly gives light to the fact that it can be used with careful tuning for data augmentation tasks too. However, analysis is going on if this helps all the models for various dataset universally but it should be considered as a good start to a new approach of data augmentation using adversarial machine learning.

\bibliographystyle{IEEEbib}
\bibliography{strings}

\end{document}